\documentclass[pre,aps,twocolumn,superscriptaddress,amsmath,showpacs,floatfix]{revtex4}
\usepackage{graphicx}
\begin{document} 
 \title{One-dimensional gas of hard needles}
 \author{Yacov Kantor}
 \email{kantor@post.tau.ac.il}
 \affiliation{Raymond and Beverly Sackler School of Physics and
 Astronomy, Tel Aviv University, Tel Aviv 69978, Israel}
 \author{Mehran Kardar}
 \affiliation{Department of Physics, Massachusetts Institute of
 Technology, Cambridge, Massachusetts 02139, USA}
 
 \date{\today}
 
\begin{abstract}
We study a one dimensional gas of needle-like objects as a testing ground 
for a formalism that relates the thermodynamic properties of ``hard" 
potentials to the probabilities for contacts between particles. Specifically, 
we use Monte Carlo methods to calculate the pressure and elasticity 
coefficient of the hard--needle gas as a function of its density. The results 
are then compared to the same quantities obtained analytically from a 
transfer matrix approach.
\end{abstract}
\widetext
 
\maketitle

 \section{Introduction}
 
 Due to the relative simplicity of the derivation of their thermodynamic properties,
 classical one-dimensional (1D) systems are frequently employed as a test bed
 of theory and methods for collective behavior in higher dimensional systems.
 For instance, the collection of ``hard spheres" on a line, sometimes referred to as 
 the Tonks gas~\cite{tonks}, has served as an initial step in the study of two-
and three-dimensional systems of hard disks/spheres.
There is indeed a general method for exact analysis of a gas of point particles interacting
in 1D  via potentials that depend only on near-neighbor separations~\cite{takanashi}.
Here, we employ such methods to study a gas of hard needle-shaped objects.
Our object is to compare the analytical results with those obtained by (Monte Carlo
implementation) of a formalism that relates thermodynamic properties (specifically
the pressure and elasticity coefficient) of the gas to the probabilities of contact
amongst the particles.

``Hard" potentials, which are either zero or $\infty$, help to illuminate the
geometrical/entropic features of a thermodynamic system.
Since there is no energy scale arising from such potentials, the temperature
$T$ appears only as a multiplicative factor in the free energy and various
other thermodynamic quantities, such as pressure and elastic coefficients.
Thus the state of the system becomes independent of $T$, and only depends
upon such features as density. 
The clarity of the geometrical perspective, combined with the simplicity of
numerical simulations, has lead to extensive studies of such systems.
In fact,  simulations with hard  potentials date back to the origins of the 
Metropolis Monte Carlo (MC) method~\cite{origMetro}, and have flourished
in the decades that have followed (see Ref.~\cite{gast} and references therein). 
A typical example of non-trivial behavior is the entropically driven first order phase 
transition from a liquid to a solid phase~\cite{hsfreezing}. 

Alignments of non-spherically-symmetric molecules lead to a diversity of phases
in {\em liquid crystals}~\cite{degennes_liquid}.
For example, in the nematic phase the molecules have no positional order (like
a liquid), while their orientations are aligned to a specific direction.
From the early stages research into liquid crystal it was realized that the {\it entropic}
part of the free energy related to non-spherical shapes of the molecules, 
can by itself explain many of the properties of such systems~\cite{ons}.  
Not surprisingly, hard potentials were frequently invoked, and even such simplifications
as infinitely thin disks~\cite{frenkel_thindisk} or rods~\cite{frenkel_thinrod} have 
provided valuable insights regarding liquid crystals.
The interplay between the rotational and translational degrees of freedom
in molecular solids~\cite{plastic} leads to elastic properties that are coupled to orientational
order. How does one compute the elastic response of such systems from first principles?

\begin{figure}
\includegraphics[width=7cm,clip]{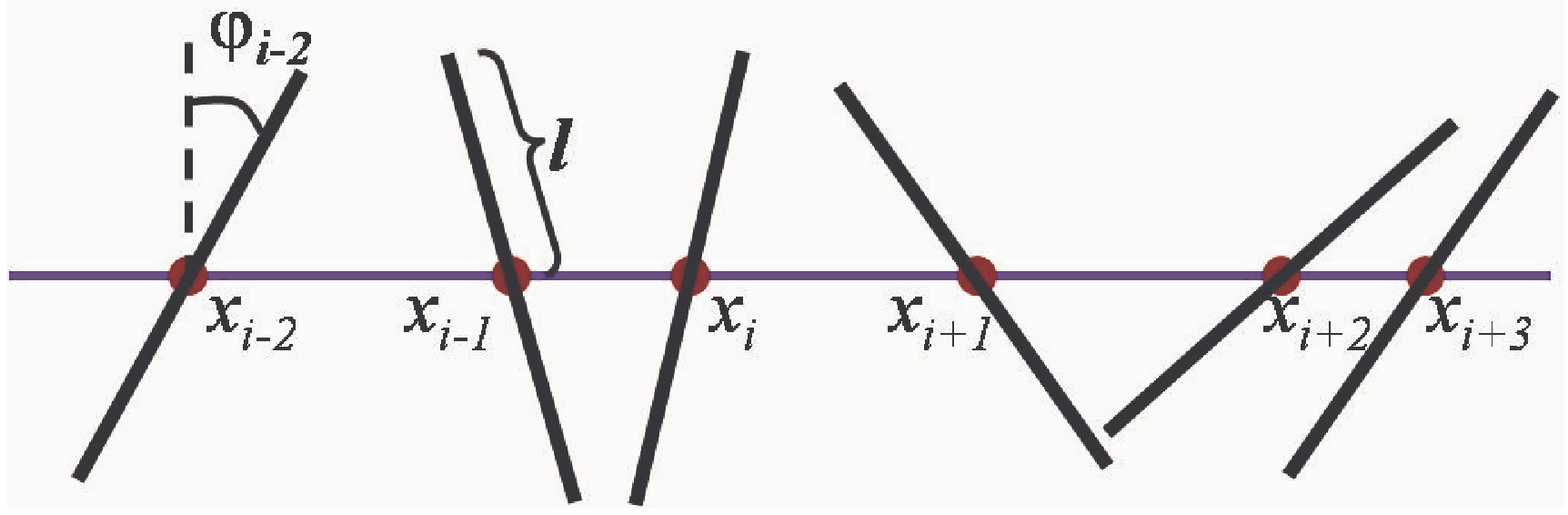}
\caption{\label{fig:1Drot} (Color online) Needle-shaped particles of length
$2\ell$  and vanishing thickness are free to rotate in two dimensions, 
with their centers  moving along a line. Particle (needle) $i$  is 
characterized by its translational position $x_i$ and orientation
angle $\phi_i$.
 }
\end{figure}

Recently a formalism enabling direct calculation of elastic
properties and stresses of a system of hard 
non-spherically-symmetric objects was developed~\cite{mk}
by extending a previously known formalism for hard spheres~\cite{fk}. 
Not surprisingly, given that the elastic response in two and higher
dimensions depends on a rank four tensor, the resulting expressions 
contain a large number of terms. Typical terms correspond to a variety 
of  possible contacts between particles and numerous components of the 
separations between them (see, e.g., Eq. 23 in Ref. \cite{mk}). 
Since these expressions are obtained after numerous mathematical 
transformations, it is advisable to subject them to independent tests.
Indeed they have been shown to reduce to the known results for isotropic 
objects, but up to now there had been no comparison with exact results 
for non-spherical particles. Here, we consider the statistical mechanics 
of a 1D system of hard needle-like particles rotating in two dimensions 
with their centers affixed to a 1D line, as depicted in Fig.~\ref{fig:1Drot}.
The  needles are not allowed to intersect, and thus act as ``hard" potentials. 
This model is a particular case of a group models considered by Lebowitz 
{\it et al.}~\cite{lebowitz} with anisotropic objects in one dimension. 
From the perspective of complexity, such systems are a slight generalization 
of the Tonks gas, yet they provide non-trivial insights  into the interplay 
of rotational and translational degrees of freedom. The model can be solved 
exactly, and thermodynamic properties, such as elastic coefficients, can be 
calculated. We compare the values obtained analytically by the transfer 
matrix method, with those from MC simulations using the expressions from 
Ref.~\cite{mk} adapted to the 1D case.

The paper is organized as follows:
The model of hard needles is introduced in Sec.~\ref{sec:model}, and we 
demonstrate how the relative orientations of neighbors leads to an 
effective hard-potential as a function of their separation. Sec.~\ref{sec:elast} 
is devoted to reviewing how elastic properties of a system can be characterized, 
and the expressions for computing elastic coefficients in 1D are presented.
The numerical difficulties associated with evaluation of various quantities 
by MC simulations are also described. In Sec.~\ref{sec:transf} we present the 
transfer matrix method for solution of the model. Details of the MC simulation 
are presented in  Sec.~\ref{sec:results}, along with comparisons to results 
obtained by transfer matrix method. Discussion and additional features
of the model are presented in Sec.~\ref{sec:discussion}.

\section{Model}\label{sec:model}

Figure~\ref{fig:1Drot} depicts a configuration of our model, consisting of
needles of length $2\ell$, with their center positions restricted to move 
on a 1D line. Needle $i$ is characterized by its position $x_i$, and orientation 
$\phi_i$ measured with respect to the normal to the line. Since orientations 
differing by $\pi$ are indistinguishable, we restrict 
$-\pi/2\le\phi_i<\pi/2$. (Such entities, called {\em directors}, frequently 
appear in the description of liquid crystals~\cite{degennes_liquid}.) 
As $\ell$ is the only microscopic length scale in the  problem, it can be 
used to construct  dimensionless parameters.  In particular, the mean distance 
between particles $a$ is made dimensionless by considering $a/\ell$, while the 
density $n$ can be replaced by $n\ell$.

\begin{figure}
\includegraphics[width=8cm]{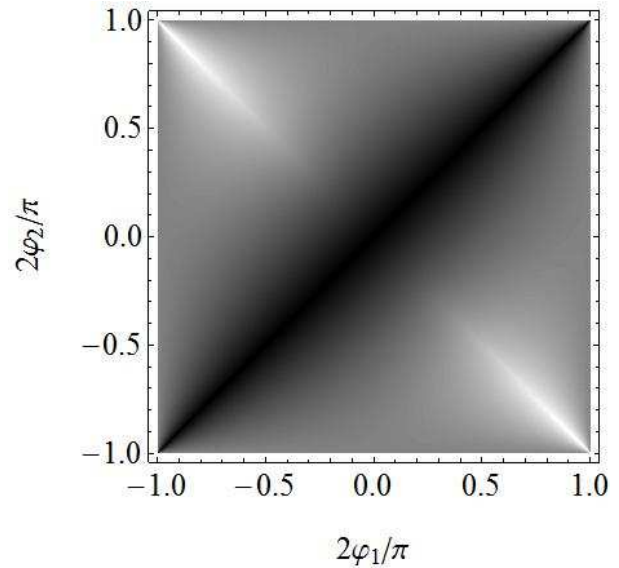}
\caption{\label{fig:contour} Gray-level representation of the
function $d_{1,2}(\phi_1,\phi_2)$ in Eq.~(\ref{eq:d}) for the dependence
of minimal distance between needles on their orientations.
The black diagonal corresponds to $d=0$ for parallel needles,
while white corresponds to $d=2$ for needles along the line.
 }
\end{figure}

The needles are not allowed to intersect but do not interact otherwise.  
Since the particles cannot cross each other, we number them (left-to-right) 
along  the 1D line, and require that this order is unchanged, i.e. 
$x_{i-1}< x_{i}$. (This convention simplifies the enumeration of the possible 
contacts between particles.) Thus the distance of closest approach between 
adjacent needles is a function of their orientations, given $\ell d$, with 
the dimensionless function
\begin{equation}\label{eq:d}
d_{i-1,i}(\phi_{i-1},\phi_{i})=
\frac{\sin|\phi_i-\phi_{i-1}|}{ \max[\cos(\phi_{i-1}),\cos(\phi_i)]}\,.
\end{equation}
This function is depicted in Fig.~\ref{fig:contour}, and varies between zero 
(when the needles are parallel, $\phi_i=\phi_{i-1}$) to 2 (with the needles 
lying on the line, $\phi_i=-\phi_{i-1}\to\pi/2$). Note that value of $d$ is 
poorly-defined at the points $(\pm\pi/2,\pm\pi/2)$, and depends on the limiting 
procedure. Analytic computations would have been considerably simplified if 
$d$ was only a function of the difference in orientation, but this is not the 
case because of the denominator in Eq.~(\ref{eq:d}).

We consider a collection of $N$ needles, either in an ensemble of fixed length 
$L$ (for MC simulations), or fixed external pressure (force) $p$ (for transfer 
matrix studies). It is convenient to impose the {\em boundary conditions} 
through the definition of the minimal distance.
For the MC simulations, we introduce fictitious particles $i=0$ an $i=N+1$.
Periodic boundary conditions on a line of length $L$ are implemented by requiring
$x_{N+1}=x_1+L$ and $\phi_{N+1}=\phi_1$, while $x_0=x_N-L$ and $\phi_0=\phi_N$. 
This extends the validity of Eq.~(\ref{eq:d}) to $i=1$ and $i=N+1$, enabling 
the treatment of all particles on equal footing.  In the fixed pressure 
ensemble, which is used in transfer matrix calculations, the orientation of 
the first (last) particle $i=1$ ($i=N$) is restricted only by its neighbor from 
the right (left), i.e. $i=2$ ($i=N-1$). The position of the both end  particles 
is arbitrary in this ensemble, with $x_1>0$ and $x_N=L$ (which is also a 
variable in this ensemble). In this case, we set $d_{0,1}=d_{N,N+1}=0$.

As explained above, adjacent particles interact via the hard potential
\begin{align}\label{eq:V}
V_{i-1,i}\equiv V(x_i-x_{i-1},\phi_{i-1},\phi_i)=\nonumber \\
\begin{cases}0, &\text{if $x_i-x_{i-1}>\ell d_{i-1,i}$,}\\
\infty, &\text{otherwise.}
\end{cases}
\end{align}
As such a potential does not have an energy scale, the temperature $T$ 
will appear only as a prefactor in the thermodynamic quantities. In 
particular, the Helmholtz free energy $F$ which is an extensive quantity 
with units of energy will have a form $F=Nk_BT h(n\ell)$, where $k_B$ is 
the Boltzmann constant. In 1D, the pressure $p$, and the elastic  coefficient 
$C$ have units of force and can be made dimensionless by considering 
$f\equiv\beta p\ell$ and $\beta C\ell$, where $\beta=1/k_BT$. The Gibbs free 
energy $G=Nk_BT g(f)$ depends only on the dimensionless pressure $f$.

\section{Elasticity of 1D}\label{sec:elast}

Shape and size deformations of objects are usually described by the
strain tensor~\cite{lanlif}. In 1D this reduces to a scalar quantity 
$\eta$ which simply relates the distorted size of the system $L'$ to its 
original size $L$ via $L'^2=L^2(1+2\eta)$~\cite{etausage}. (While this 
definition is slightly awkward in 1D, the use of squared distances between 
points is convenient because in higher dimensions it clearly separates 
trivial changes in geometry caused by rotations, and real deformations.)
In 1D, for small $\eta$, the Helmholtz free energy  can be expanded as 
\begin{equation}\label{eq:elast_def}
\frac{F(\eta)}{L}=\frac{F(0)}{L}-p\eta+\frac{1}{2}C\eta^2+\dots\,,
\end{equation}
where $C$ is the elastic coefficient of the body. (Note, that the free
energy on the left hand side (l.h.s.) of the equation is divided by 
the {\em undistorted} size of the system.) Consequently, $p$ and $C$ can 
be calculated from the first and second derivatives of $F$ with respect to 
$\eta$ at fixed $T$.

While  the elastic properties are more naturally obtained from the Helmholtz 
free energy, we will also use the Gibbs free energy $G=F+pL$, in which the 
pressure is the (imposed) variable~\cite{kardar}. The system size $L$, or the 
mean inter-particle distance is then obtained from 
$a=\left.\partial (G/N)/\partial p\right|_T$, or in terms of dimensionless 
variables 
\begin{equation}\label{eq:a_from_G}
\frac{a}{\ell}=\left.\frac{\partial (\beta G/N)}{\partial f}\right|_T\, .
\end{equation}
Similarly, $C=-\left. a/\left(\frac{\partial a}{\partial p}\right|_T\right)+p$,
and
\begin{equation}\label{eq:C_from_G}
\beta \ell C=-\frac{a}{\left.\frac{\partial a}{\partial f}\right|_T}+f \, .
\end{equation}

Squire {\it et al.}~\cite{shh} developed a formalism for a direct calculation
of elastic parameters from the correlation functions of particles. 
In this approach stress (pressure) and elastic moduli are related to thermal 
averages of products of various inter-particle forces and separations.
This formalism was extended to hard potentials in Refs.~\cite{fk,mk}.
Since for hard potentials the forces vanish except when the particles 
touch, the results depend on various contact probabilities. In two and three 
dimensions the stress and the elastic constants are tensors, and the 
expressions involve averages over a variety of components. These results 
simplify in 1D, and in particular, the expression for stress
[Eq.~(22) in Ref.~\cite{mk}] can be considerably simplified: 
If we denote the separation between two adjacent needles by
$s_i=x_{i+1}-x_i$, they are in contact if the argument of
$\Delta_i=\delta[s_i-\ell d_{i,i+1}(\phi_i,\phi_{i+1})]$ vanishes. The
dimensionless pressure then $f$ becomes
\begin{equation}\label{eq:f}
f=n\ell\left(1+\frac{1}{N}\sum_i\langle s_i\Delta_i\rangle\right).
\end{equation}
The first term in this expression is simply the pressure of the ideal gas,
while the second term can be easily recognized as the mean value of the 
product of the inter-particle separation and force, as appears in the virial theorem~\cite{virtheorem}. 
To evaluate Eq.~(\ref{eq:f}), we need the probability that two particles
($i$ and $i+1$) with specified orientations ($\phi_i$ and $\phi_{i+1}$) touch 
each other.

Similarly, the elastic coefficient $C$ can be expressed as
\begin{eqnarray}\label{eq:C}
\beta C \ell=n\ell\left[2+\frac{3}{N}\sum_i\langle s_i\Delta_i\rangle
+\frac{1}{N}\left(\sum_i\langle s_i\Delta_i\rangle\right)^2\right. \\ \nonumber
\left.-\frac{2}{N}\sum_{i<j}\langle s_is_j\Delta_i\Delta_j \rangle
-\frac{1}{2N}\sum_i \langle (s_i^2+s_{i+1}^2)\Delta_i\Delta_{i+1}\rangle\right].
\end{eqnarray}
The last two sums in the right hand side (r.h.s.) of Eq.~(\ref{eq:C}) involve 
averages of products of $\Delta$s, i.e. they require knowledge of the {\em joint 
probability density} of two simultaneous contacts. The last sum involves cases 
when three particles $i$, $i+1$ and $i+2$ touch each other, while the
preceding sum depends also on cases when two independent pairs are in contact, 
i.e. particle $i$ touches $i+1$, and a different particle $j$ ($>i+1$)
touches $j+1$.  The l.h.s. of Eq.~(\ref{eq:C}) is an intensive quantity, 
while the third and the fourth terms on its r.h.s. contain $O(N^2)$ terms. 
However, most of the terms appearing in these two sums can be grouped in pairs 
$\langle s_i\Delta_i\rangle\langle s_j\Delta_j\rangle-
\langle s_is_j\Delta_i\Delta_j\rangle$, which decay to zero
when the distance between the pairs of particles exceeds the correlation length. 
All the averages appearing in Eqs.~(\ref{eq:f}) and (\ref{eq:C})
can be calculated in MC simulations.

\section{Transfer matrix approach}\label{sec:transf}

The partition functions of 1D models with short range interactions can be 
found analytically using a transfer matrix method~\cite{baxter,casey,kardar}. 
It convenient to consider the isobaric ensemble with fixed external pressure
(force) $p$, such that (the configurational part of) the partition function 
is given by
\begin{equation}\label{eq:Zdef}
Z_G=\int\prod_{i=1}^Ndx_i\prod_{i=1}^Nd\phi_i
{\rm e}^{-\beta\sum_{i=1}^NV_{i-1,i}-\beta p x_N}.
\end{equation}
Since $x_N=\sum_i s_i$, we can change variables and perform integrations 
over the separations $s_i$ between adjacent particles. For the hard potential 
given by Eq.~(\ref{eq:V}),  this leads to
\begin{align}\label{eq:ZdefRes}
Z_G&=(\beta p)^{-N}\int\prod_{i=1}^N\left(d\phi_i{\rm e}^{-\beta p \ell d_{i-1,i}(\phi_{i-1},\phi_i)}\right) \nonumber\\
&=(\beta p)^{-N}\int\prod_{i=1}^N\left(d\phi_i D_{i-1,i}(f;\phi_{i-1},\phi_i)
\right),
\end{align}
where
\begin{equation}
D_{i-1,i}(f;\phi_{i-1},\phi_i)\equiv  {\rm e}^{-f d_{i-1,i}(\phi_{i-1},\phi_i)}.
\end{equation}
(According to the definition of $d$, all $D$ are identical, except for
 $D_{0,1}=D_{N,N+1}\equiv 1$ at the boundaries, as explained in the 
 Sec.~\ref{sec:model}.)

The expression in Eq.~(\ref{eq:d}) is too complicated for the integrals in 
Eq.~(\ref{eq:ZdefRes}) to be performed analytically. Nevertheless, multiple 
integrals of this kind can be easily performed numerically to any desired 
accuracy. We can subdivide the range of the angular integration into $M$ 
equal segments, by setting $\phi_k=-\pi/2+\frac{\pi}{M}k$,  with 
$k=0,1,\cdots, M-1$. This replaces the function $D$ by an $M\times M$ matrix, 
and the integrals in Eq.~(\ref{eq:ZdefRes}) are replaced by matrix products. 
The partition function then becomes
\begin{equation}
Z_G=(\beta p)^{-N} (\pi/M)^N \tilde{v}D^{N-1}v,
\end{equation}
where $v$ is a column vector with all of its elements equal to 1. 
Repeated multiplications can be performed numerically, first multiplying 
$D$ by itself, then multiplying the resulting matrix by itself, etc..
After a total of $K$ such iterations we arrive at $D^{N}$, with $N=2^K+1$.
The exponential dependence on $K$ allows us to achieve very large values 
of $N$, in practice we used $K=20$ in our simulations. For moderate 
pressures,  the discretization of the angle $\phi$ has little influence on 
the result once $M$ exceeds 10, and we report results for $M=512$. 
(It should be noted, that the same results can also be obtained by
numerically finding the largest eigenvalue of $D$. However, in our
case this alternative provides no numerical advantage.) From the 
numerical value of $Z_G$, we then obtain the Gibbs free energy

\begin{figure}
\includegraphics[height=5.5cm]{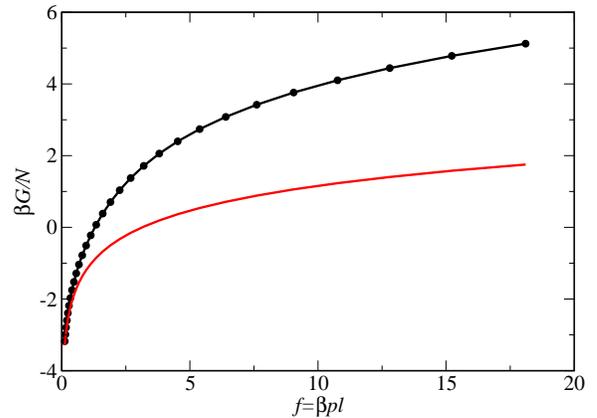}
\caption{\label{fig:free_energy} (Color online) The upper curve 
depicts the Gibbs free energy per  particle (made dimensionless by 
multiplying by $\beta$) as a function of the dimensionless pressure
$f=\beta p \ell$.  The lower curve shows, for comparison, the 
same quantity for non-interacting needles; the curves begin to 
separate when $f$ is larger than about one.}
\end{figure}

Figure~\ref{fig:free_energy} depicts the scaled Gibbs free energy calculated 
by this numerical procedure. For non-interacting needles the partition
function is $Z_0=(\pi/\beta p)^{-N}$, and the corresponding 
$\beta G_0/N=\ln(\beta p/\pi)$ is indicated by the lower line in  the figure. 
(Both curves exclude the trivial contribution due to kinetic energy.)

\section{Simulations and  results}\label{sec:results}

A Monte Carlo procedure was used to evaluate the pressure and elastic 
coefficient of the system of hard needles. We simulated $N=128$ particles 
with periodic boundary conditions. Correlations between the needles for
small and moderate densities do not persist past a few neighbors,
and our choice of $N$ caused no perceptible finite size effects. An 
elementary MC move consists of randomly choosing a needle, randomly 
deciding whether to displace or to rotate it, and attempting to perform 
such a move. The move is accepted if in the new position, or with new 
orientation, the needle does not overlap with neighboring needles. The 
particles are sequentially ordered, and a position change is rejected if 
it changes this ordering. The attempted moves are uniformly distributed 
over an interval, whose width is chosen to be as large as possible, while 
maintaining an acceptance rate larger than 50\%. The varying size of the 
interval implies a diffusion constant for each particle that decreases with 
increasing density. A single MC time unit consists of $2N$ attempts to move 
or rotate particles. The relaxation time of the system is proportional to 
$L^2$, and inversely proportional to the diffusion constant and elastic 
coefficient.  (The latter increases with increasing density.) Our choice of 
elementary step ensured that, within the examined range of densities, the 
relaxation time was approximately constant and remained of order $N^2$. This 
was verified by directly measuring several autocorrelation functions.  For 
every density $n$ the simulation time was $5\times 10^5N^2$. Such long times
are required to ensure high accuracy of measured contact probabilities, 
as explained below.

The presence of the Dirac $\delta$-function in the definition of $\Delta_i$
necessitates delicate handling. Both Eq.~(\ref{eq:f}) and Eq.~(\ref{eq:C})
require measuring separation $s_i$ between the adjacent needles at the moment 
of contact. Such events have zero probability, and the formulas really involve 
probability {\em densities}. The latter can be evaluated by examining the 
probability that the two needles are within  $\epsilon_1$, and divide the result 
by $\epsilon_1$. Of course, the number of such near collisions decreases with 
decreasing $\epsilon_1$, and the statistical error increases. The situation is 
even worse for terms of type $\langle s_is_j\Delta_i\Delta_j\rangle$, where 
{\em two} simultaneous contacts are supposed to appear. One may define two near 
collision events by considering intervals of size $\epsilon_1$ and $\epsilon_2$. 
The opposing requirements of having $\epsilon_i\rightarrow 0$ (for accurate 
calculation of probability densities), and large $\epsilon_i$ (to ensure 
statistical accuracy) can be partially reconciled by considering each argument 
of a $\delta$-function being in the range $[m\epsilon,(m+1)\epsilon)$, with 
$m=0,1,\cdots,M$. We used $M=10$, and $\epsilon=0.002$ (0.01) for high (low) 
particle density simulations. With 11 data points for single contact terms, and 
$11^2$ points for two-contact terms, we could view the results as a function of 
one variable $m_1$, or two variables $m_1$ and $m_2$, and extrapolate the 
results to the ``exact contact" limit. The  accuracy and practicality of such a 
procedure has been demonstrated in Ref.~\cite{fk}. The total simulation time was 
determined by requirement of having sufficient number of terms in each ``bin" of 
the statistical procedure explained above. The total simulation time was
dictated by the need to have a very accurate estimate of the fourth term on the
r.h.s. of Eq.~(\ref{eq:C}).

\begin{figure}
\includegraphics[height=5.5cm]{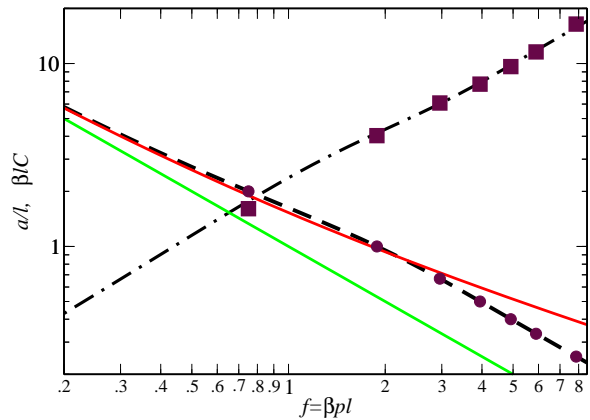}
\caption{\label{fig:mean_dist} (Color online) The mean distance between
particles $a$ (in units of $\ell$), as a function of dimensionless force $f$,
for an ideal gas (lower continuous line) as compared with the virial expansion 
truncated at the second term (upper continuous line),
and with the exact transfer matrix result (dashed line).
The dashed-dotted line is the transfer matrix result for the elasticity
coefficient $C$. Solid circles represent the relation between $a$ and $f$
from the MC simulations. Solid squares represent
the MC results for the elastic coefficient. 
}
\end{figure}

Since the MC simulation is performed in the ensemble of fixed length, the density or
mean inter-particle distance $a$ are given, while the dimensionless pressure $f$
and the dimensionless elastic modulus $\beta\ell C$ are calculated. 
The full circles in Fig.~\ref{fig:mean_dist} depict the calculated dependence of $f$ 
(horizontal axis) on $a$ (vertical axis). 
The error bars on $f$ are negligible, since Eq.~(\ref{eq:f}) includes only single pair contacts, 
and the large statistics as well as small $\epsilon_1$ ensures very high accuracy. 
This result is compared with the relation between $f$ and $a$ obtained from the 
Gibbs free energy  via Eq.~(\ref{eq:a_from_G}) by taking the numerical derivative 
of $G$ calculated by the transfer matrix method. 
The latter is depicted by the dashed line. Excellent agreement is obtained between
the results from these two methods.

The solid squares in Fig.~\ref{fig:mean_dist} depict the MC results for the 
dimensionless elastic coefficient  $\beta\ell C$ (vertical axis), 
as a  function of the dimensionless pressure $f$ (horizontal axis). 
Since in the MC procedure $f$ is itself a computed quantity, there are now also
horizontal error bars, which are negligible as explained in the previous paragraph.
The accuracy of $C$, however, is much lower and depends on both statistical errors,
and systematic errors from extrapolation to the true contact probability densities. 
We chose the values of $\epsilon_i$, and the simulation time in 
such a way that both errors were of the same order. We estimate that the vertical 
error bars are approximately the size of the symbol for the leftmost point, and 
decrease to half the symbol size for the rightmost point. The dashed-dotted line 
depicts the same relation obtained from $G$ by using Eq.~(\ref{eq:C_from_G}) and 
the transfer matrix calculations. The results from both approaches coincide within
the estimated errors.

\section{Discussion}\label{sec:discussion}

The good agreement between the results from MC simulations, based
on contact probabilities, and those from the transfer matrix method,
support the validity of the expressions reproduced in Eqs.~(\ref{eq:f}),~(\ref{eq:C})
for the pressure and elastic moduli of hard potentials. While limited to 1D, 
this is the first direct comparison between formulae derived in Ref.~\cite{mk}, 
and an exact alternative approach.

We conclude by pointing out an interesting feature of the hard needle system:
Both at small and large densities, the pressure and elastic coefficient are
related by the simple expression $C=2p$, while the behavior at intermediate
densities is more complicated. For low pressure (density) the system behaves 
as an ideal gas with $a=(\beta p)^{-1}$, and substituting in 
Eq.~(\ref{eq:C_from_G}) immediately yields $C=2p$ in this limit. Interestingly, 
as discussed by Lebowitz {\it et al.}~\cite{lebowitz}, the relation between 
density and pressure also simplifies at very high pressure (density). In this 
limit the angular integrations are themselves constrained by pressure, and a 
Gaussian approximation leads to additional powers of $\beta p$ in the Gibbs 
partition function. This in turn leads to a density $n=a^{-1}=2\beta p$, i.e. 
the same functional dependence as an ideal gas but with a factor of 2.
Inserting this limiting behavior into Eq.~(\ref{eq:C_from_G}) again leads to 
 $C=2p$ as in the ideal gas limit.

The lower solid line if Fig.~\ref{fig:mean_dist} depicts the the dependence of 
$a$ on $f$ for an ideal gas. The curve deviates from ideal behavior for values
of $f$ larger than about 1. At higher densities (and pressures) we can improve 
upon the ideal gas behavior by using a virial expansion. From the form of the 
interaction we compute a second virial coefficient of $B_2=8/\pi^2$. As 
indicated by the upper solid line in Fig.~\ref{fig:mean_dist}, inclusion of the 
second virial coefficient provides a good approximation for $f$ up to 3. Clearly
there is no no simple relation between $C$ and $f$ in this intermediate region.

The focus of this paper was to use the model of hard needles to validate the 
relation between elastic moduli and contact probabilities for the exactly
solvable model of hard needles. However, the model itself has some interesting
features, which will be explored elsewhere~\cite{kk_tbp}. In particular the
simplified behavior alluded above in the high density limit is related to an
incipient critical point. The nature and universality of this criticality is 
related to the shapes of the hard objects (in this case, needles). 

 \begin{acknowledgments}
This work was supported by the Israel Science Foundation Grant No. 193/05 (Y.K.) 
and by the National Science Foundation Grant No. DMR-08-03315 (M.K.) Part of 
this work was carried out at the Kavli Institute for Theoretical Physics, with 
support from the National Science Foundation Grant No. PHY05-51164.
 \end{acknowledgments}

 \end{document}